\documentclass[apj]{emulateapj}
\usepackage{apjfonts}
\def\lax {\ifmmode{_<\atop^{\sim}}\else{${_<\atop^{\sim}}$}\fi}  
\def\gax {\ifmmode{_>\atop^{\sim}}\else{${_>\atop^{\sim}}$}\fi}  
\def\gtorder{\mathrel{\raise.3ex\hbox{$>$}\mkern-14mu
             \lower0.6ex\hbox{$\sim$}}}
 
\begin{document}

\title{Correlations between X-ray Spectral and Timing Characteristics 
in Cyg~X-2}

\author{Lev Titarchuk\altaffilmark{1}, Sergey Kuznetsov\altaffilmark{2}
and Nikolai Shaposhnikov\altaffilmark{3}}

\shorttitle{X-RAY SPECTRUM AND QPO IN CYG~X-2}

\shortauthors{TITARCHUK, KUZNETSOV \& SHAPOSHNIKOV}

\journalinfo{The Astrophysical Journal, 667: 000-000, 2007, October 1}

\slugcomment{accepted 2007 June 8}

\altaffiltext{1}{George Mason University/Center for Earth Observing
and Space Research, Fairfax, VA 22030; US Naval Research Laboratory,
Code 7655, Washington, DC 20375; NASA GSFC, code 661, 
Greenbelt MD 20771, USA; email:lev@milkyway.gsfc.nasa.gov}
\altaffiltext{2}{IGPP, University of California, Riverside, CA 92521, USA;
Space Research Institute of Russian Academy of Sciences,
Profsoyuznaya 84/32, Moscow 117997, Russia}
\altaffiltext{3}{ NASA GSFC/USRA, Astrophysics Science
 Division, Greenbelt MD 20771; nikolai@milkyway.gsfc.nasa.gov}

\begin{abstract}
Correlations between the quasi-periodic oscillations (QPOs) and the
spectral power-law index have been reported for a number of black hole
candidate sources and for four neutron star (NS) sources, 4U~0614+09,
4U~1608-52, 4U~1728-34 and Sco~X-1.  An examination of QPO frequencies
and index relationship in Cyg~X-2 is reported herein.  The RXTE
spectrum of Cyg~X-2 can be adequately represented by a simple
two-component model of Compton up-scattering with a soft photon
electron temperature of about 0.7 keV and an iron K-line.  Inferred
spectral power-law index shows correlation with the low QPO
frequencies.  We find that the Thomson optical depth of the Compton
cloud (CC) $\tau$, in framework of spherical geometry, is in the range
of $\sim 4-6$, which is consistent with the neutron star's surface
being obscured.  The NS high frequency pulsations are presumably
suppressed as a result of photon scattering off CC electrons because
of such high values of $\tau$.  We also point out a number of
similarities in terms timing (presence of low and high frequency QPOs)
and spectral (high CC optical depth and low CC plasma temperature)
appearances between Cyg~X-2 and Sco~X-1.
\end{abstract}


\section{Introduction}

Cyg~X-2 is a typical low mass X-ray binary (LMXB) Z-source (see van
der Klis 2000 for a review) exhibiting features which can be
attributed to Z-sources: Z-shape color-color diagram (CCD), 15-60 Hz
quasi-periodic oscillation (QPO) frequencies for horizontal branch
oscillations (HBO) and 5-20 Hz frequencies for normal-flaring branches
(N/FBO).  Large numbers of strong HBO and N/FBO frequencies have been
found in power density spectrum (PDS) of this source.  One of the
particular features of Cyg~X-2 as a Z-source is that the Z-track
undergoes secular changes (evolution) on the time-scale of days
\citep{kuu96}. There is still no consensus in the community on the 
nature of these changes.

When considering the components of the X-ray spectrum rather than
X-ray colors \cite{kaa98} have demonstrated the clear correlation
between kHz QPOs and spectral shape in neutron star (NS) LMXBs.  When
fitting the X-ray spectra of 4U~0614+09 and 4U~1608-52 with simple
power laws, \cite{kaa98} observed a correlation between the index of
the power law and the QPO frequency, the larger the index, the higher
the QPO frequency.  On the other hand \cite{ford97} have shown that
there is a one-to-one correlation between the flux of the blackbody
(BB) component of the X-ray spectrum and the QPO frequency in
4U~0614+09, it worth noting that no such correlation exists between
the total X-ray flux and the QPO frequency.  One can infer the disk
mass accretion rate $\dot M_d$ using the blackbody flux.  Combination
of these QPO-index and QPO-BB flux correlations imply that the
spectral index increases when $\dot M_d$ increases.

\cite{barret} reviewed the broad band (0.1-200 keV) spectral and timing
observations of LMXBs performed by Beppo-SAX and RXTE and discussed
all of these aforementioned correlations in detail.

\cite{tit05},
hereafter TS05, suggested that black hole (BH) and NS binaries could
be observationally distinguished by the correlation between the
spectral indices and QPO low frequencies in the binary X-ray spectra.
They reported that there was a correlation between the spectral
indices and QPO frequencies in the NS binary (the Atoll-source)
4U~1728-34. TS05 implemented a thorough analysis of spectral and
temporal properties of 4U~1728-34 using RXTE data.  They studied a
spectral evolution of this source from hard spectra at low
luminosities to soft spectra at high luminosities. They found that the
hardest spectra are described by the sum of Compton upscattering of
the disk and NS surface soft photons. These spectra are very similar
to the hard-state spectra of black hole candidates (BHCs) but they are
softer ($\Gamma\sim 1.6\pm 0.1$ for BHC, $\Gamma\sim 2.1 \pm 0.1$ for
NS) as expected from the theory [see, for example, \cite{tf04}].  On
the other hand the high luminosity soft state spectrum of 4U~1728-34
consists of the sum of two blackbody-like components with color
temperatures of about 1 keV and 2.2 keV respectively. The softer
blackbody component is presumably related to the disk emission as the
harder one is related to the NS compact photosphere emission.

Recently \citet{pai06}, hereafter P06, confirmed this spectral
evolution picture of NS sources using a sample of twelve NS bright
low-mass X-ray binaries observed by INTEGRAL [high-energy ($>20$ keV)
observations].  Their sample comprises the six Galactic Z-sources and
six Atoll-sources, four of which are bright GX bulge sources while two
(H~1750-440 and H~1608-55) are weaker in the 2-10 keV range. Comparing
their results with those obtained by \citet{fal06} on 4U~1728-34, P06
were able to identify four main spectral states for NS LMXBs (see Fig.
4 in P06): very soft state (e.g. GX~3+1), intermediate state (e.g.,
GX~5-1), hard/power-law (PL) state (H~1750-440 and H~1608-55) and
low/hard state (4U~1728-34).

\begin{figure*}[tb]
\epsscale{1.0}
\includegraphics[width=3.4in,height=3.5in,angle=0]{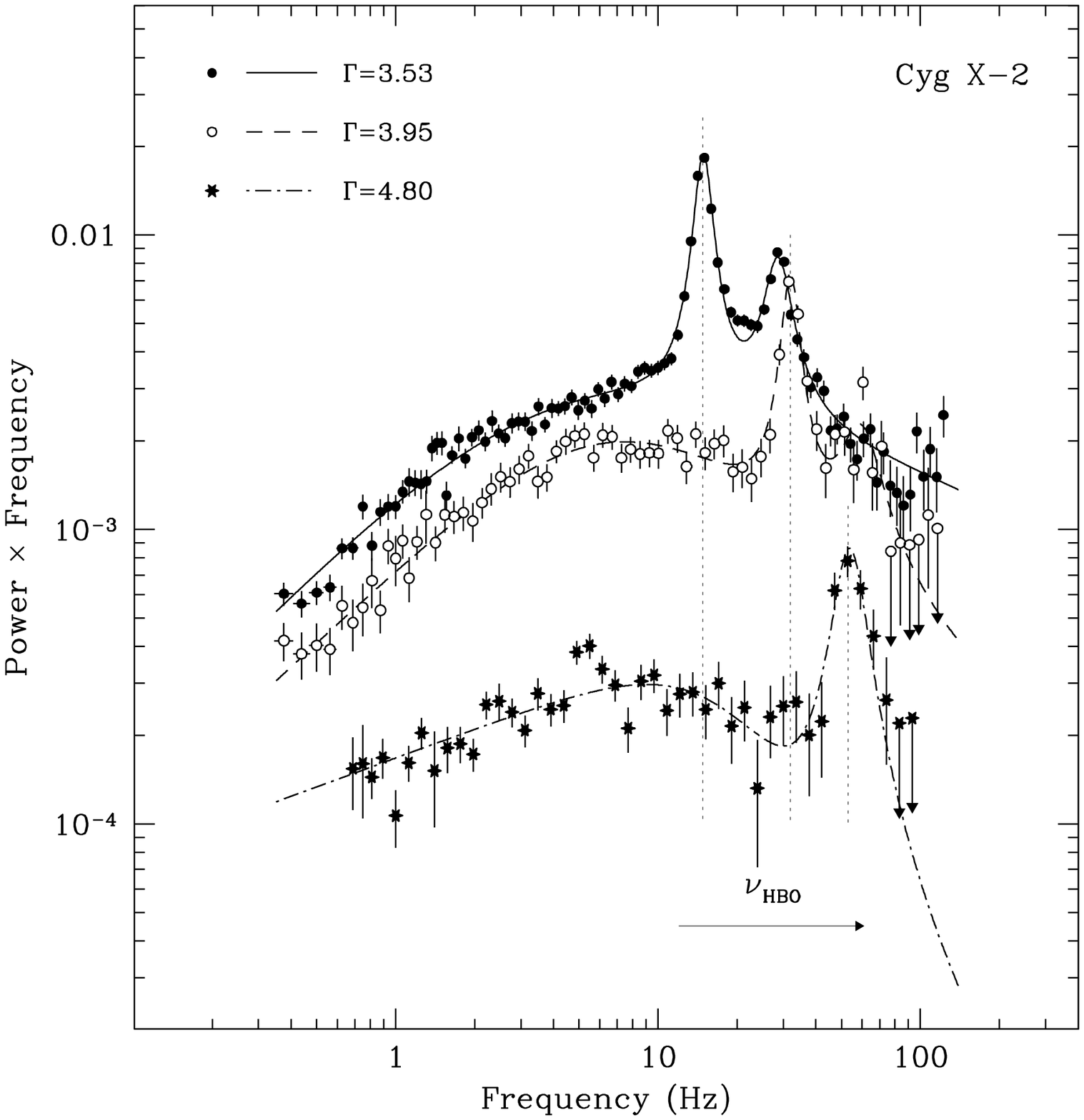}
\includegraphics[width=3.4in,height=3.5in,angle=0]{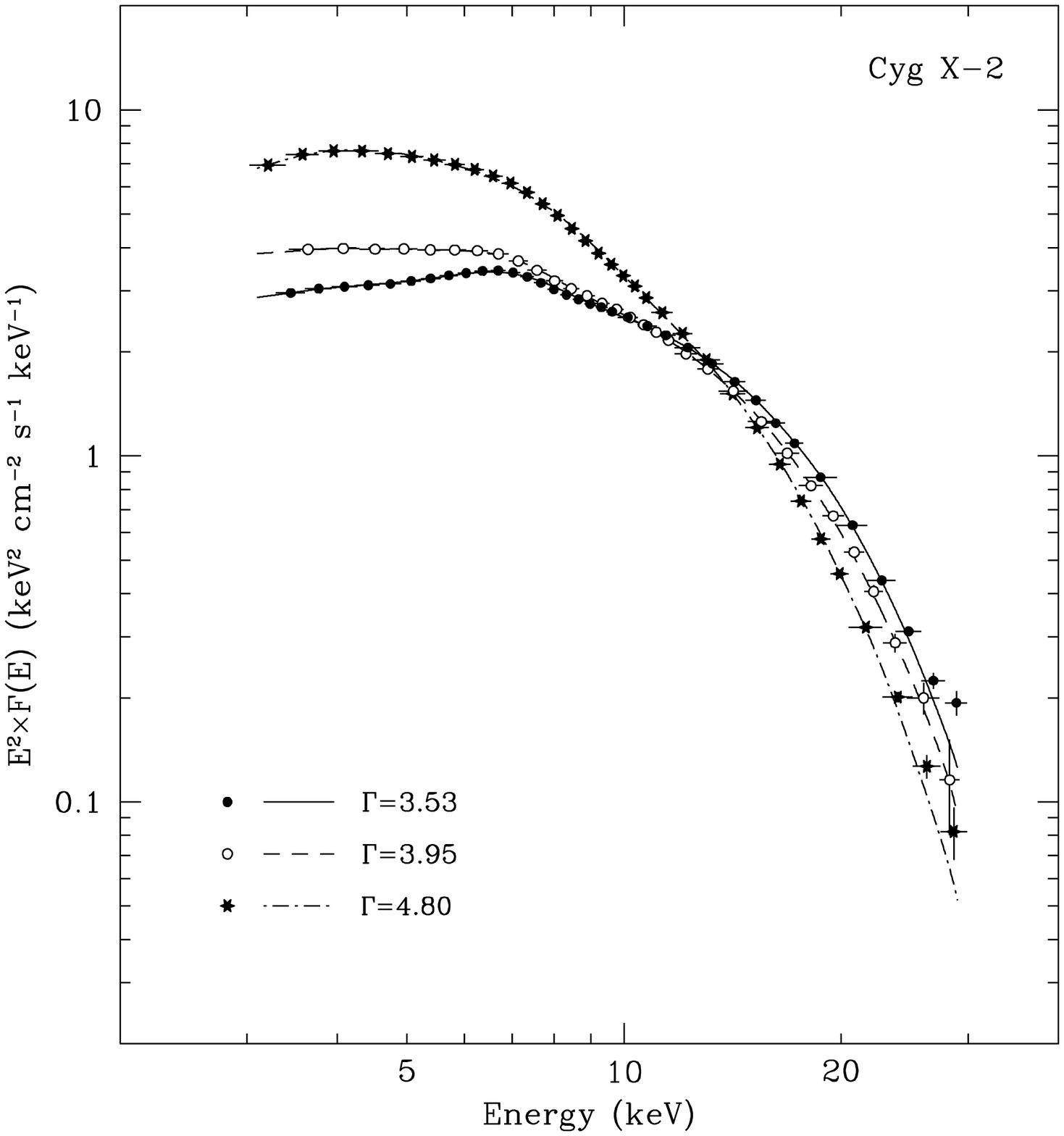}
\caption{Evolution of the power density spectra   of Z-source Cyg~X-2 
and photon spectra.  PDSs (left panel) are presented as a of
$\nu\times P(\nu)$ diagram (rms/mean)$^2$.  Photon spectra (right
panel) are presented as a $E^2\times F(E)$ diagram
(keV$^2$cm$^{-2}$s$^{-1}$keV$^{-1}$).  Power and photon spectra are
constructed for the same observational datasets observed with {\it
RXTE/PCA} on March 27, 1996 (filled circles), July 15, 1998 (starred
data points) and September 30, 2002 (open circles).  Best-fit models
of $\nu\times P(\nu)$ and $E^2\times F(E)$ continua are shown by the
curves solid curve, dashed curve and dash-and-dotted curve
correspondingly.  Strong HBO peak of horizontal branch oscillations is
a remarkable feature of Cyg~X-2 PDS and varies in wide range from from
14~Hz to 60~Hz.  Best-fit Comptonization models (COMPTT) of photons
spectra are related to the Green's function index $\Gamma$ of
$\sim3.5$, $\sim4.0$ and $\sim4.8$.  }
\label{PDSvsSpectra}
\end{figure*}

In P06 the low/hard state spectra are described by thermal
Comptonization with plasma temperature $kT \sim$ 30 keV and Thomson
optical depth $\tau \sim$ 1. The hard/power-law (PL) state spectra, as
suggested by the name, show simple PL-like emission.  The intermediate
state spectra are characterized by the presence of a thermal
Comptonization component with spectral parameters similar to those of
the soft state plus additional PL-like emission above $\sim$ 30
keV. The soft state spectra are described by a single thermal
Comptonization component with low $kT$ and high $\tau$.

The power-law state was also found by Piraino et al. (1995), hereafter
PSFK, in 4U~0614+09.  The steep extended power laws, which indices
were about $2.4$, were interpreted by PSFK as thermal Comptonization
of low energy photons on electrons having a very high temperature
$kT_e$ greater than 220 keV and small optical depth $\tau$ less than
0.2.  However one should realize that the real information in the
4U~0614+09 data is the values of the power-law indices about 2.4 but
small optical depth and high electron temperature are a matter of
interpretation in terms of the thermal Comptonization model (see \S 6
for more discussion of this small optical depth interpretation and its
relation to non-detection of pulsations from 4U~0614+09).

It is important to emphasize that some of these sources do not undergo
the spectral transition from the hard to soft states through the
intermediate state but they only evolve within a particular
state. Using RXTE data \citet{BTK}, hereafter BTK07, have found that
Sco~X-1 is always in the high luminosity soft state for which the
photon index $\Gamma$ is between 3.3 and 4.1.  The photon spectral
model is described by the thermal Comptonization component with
low-$kT$, ranging from 2.5 to 3 keV and high-$\tau$ with values at
about $5.1-6$ along with a strong 6.4 keV iron line.  These results
suggest that cooling of Compton cloud (CC) is dictated by the strong
soft emissions coming from the disk and from the neutron star. The
photons and plasma are almost in a thermodynamic equilibrium with each
other.  The CC plasma temperature is very close to the color
temperature of the blackbody-like emission of the NS compact
photosphere \citep[see e.g.][]{tit94a}.  Furthermore BTK07 demonstrate
the distinct presence of a correlation between QPOs and photon index
in Sco~X-1 similar to the previously reported correlation in
4U~1728-34. This observed correlation of kHz frequencies with photon
index in Sco~X-1 (BTK07, Fig. 3) leads to the conclusion that the CC
contracts when the source evolves to the softer states.
 
\citet{hvdk89} noted that there is a similar correlation between 
the fast time variability and the source position along the track
described in the color diagram (its color) for both high and low
luminosity sources. In this paper we suggest that in high-luminosity
LMXBs like Cyg~X-2 and Sco~X-1 {\it index-QPO correlations} should be
similar because the physical processes producing the power and photon
spectra are similar in these sources.

\begin{deluxetable*}{llllllll}[t]
\tablewidth{0pt}
\tabletypesize{\footnotesize}
\tablecaption{Results for selected RXTE observations}
\tablehead{
\colhead{Observation ID} &
\colhead{T Start} &
\colhead{T Obs.} &
\colhead{$\nu_{L}$} &
\colhead{$kT$} &
\colhead{$\tau$} & 
\colhead{$\Gamma$} & 
\colhead{$\chi^2_{red}$ ($N_{dof}$)}\\
\colhead{(orbit)} &
\colhead{(MJD)} &
\colhead{(ks)} &
\colhead{(Hz)} &
\colhead{(keV)} &
\colhead{ ~ } &
\colhead{ ~ } &
\colhead{ ~ }}
\startdata
10066-01-01-00.4 ($\bullet$)\tablenotemark{a} & 50169.6315 & 3.0 & 
$14.83\pm0.03$& $2.98\pm 0.01$ & $5.68\pm 0.01$ &
$3.53\pm 0.01$ & 0.96 (52) \\
70104-02-02-00.1 ($\circ$)\tablenotemark{a} & 52547.8956 & 1.4 &
$31.90\pm 0.18$ &  $2.94\pm0.03$& $5.04\pm 0.06$ & $3.95\pm 0.05$ & 
0.64 (46) \\
30046-01-01-00.4 ($\ast$)\tablenotemark{a} & 51009.6857 & 3.3 & 
$53.09\pm 0.97$ & $2.71\pm0.01$& $4.23\pm 0.01$ & $4.80\pm 0.01$ &
1.03 (52)\\
\enddata
\tablenotetext{a}{These symbols mark the corresponding
data points on Figs. \ref{PDSvsSpectra}, \ref{kT_tau} and
\ref{index_QPO} }
\end{deluxetable*}

A phenomenological classification of the aperiodic and quasi-periodic
variability in LMXBs has been already given in the literature, and it
is also shown that these features behaves in a similar way in Z- and
atoll-sources and perhaps also in black hole candidates [see
e.g. \citet{wk99}; \citet{pbk99}].

%%
%%\citet{tok} applied the transition layer (TL) model [see \cite{tlm98}, 
%%\cite{ot99}, \cite{to99}] to Sco~X-1's power spectra. They offered a scheme 
%%which allows the classification of the power spectral features if the
%%twin kHz QPO frequencies are observed along with the low QPO
%%frequencies.
%%

Titarchuk, Osherovich \& Kuznetsov (1999) applied the transition layer
(TL) model [see Titarchuk, Lapidus \& Muslimov (1998), Osherovich \&
Titarchuk (1999), Titarchuk \& Osherovich (1999)] to Sco~X-1's power
spectra. They offered a scheme which allows the classification of the
power spectral features if the twin kHz QPO frequencies are observed
along with the low QPO frequencies.  In this work we also adopt this
classification. In section \ref{PDS} we show that the power spectra of
Sco~X-1 and Cyg X-2 are similar.  Break frequency ($b$), horizontal
branch oscillation (HBO) frequency harmonics ($L$, $2L$), lower kHz
(Keplerian) ($K$) and higher (hybrid) ($h$) frequencies are are
clearly seen in the Cyg~X-2 PDSs.  The lowest peaks in Sco~X-1 and
Cyg~X-2 PDSs are identified in the TL model as a frequency of
magneto-acoustic (viscous) oscillations ($\nu_{V}$).

In Cyg~X-2 (similar to Sco~X-1) we find that the high/soft state
spectra (photon spectral index) varies when the low QPO frequencies
change at least by factor of 2.  Description of observations, data
reduction and analysis are presented in \S 2.  Description of Cyg~X-2
power spectra and QPO identification are presented in \S 3. Details of
our spectral fitting are presented in
\S 4 and we discuss a correlation of low QPO frequency vs photon index 
revealed in Cyg~X-2 in \S 5.  Discussion and conclusions follow in \S
6 and \S 7.

\section{Observations, Data Reduction and Analysis}
\label{observ}
In the present work we use Cyg~X-2 data collected by RXTE/PCA
\citep{jah96} during following cycles of pointing observations: 10063,
10065, 10066, 10067, 20053, 30046, 30418, 40017, 40019, 60417, 70014,
70016, 70104.  Our data set comprises 760 ksec total on-source time.
For our analysis we extract PDS and energy spectrum for each RXTE
orbit revolution.

To construct PDSs we used observational data with a resolution of
$\sim 122 \mu$s (and better) from all 249 PCA energy channels. We
combined some of the observational data that were not presented in a
single format for all 249 channels. Double event data were also used
and merged when they are available.  Among the observations of
Cyg~X-2, all five proportional counters were not always switched on to
record events. If the operating condition of one of the counters
changed during a continuous observation, then the time interval during
which the total count rate changed abruptly was excluded from our
analysis.  Note duration of the a continuous observation did not
exceed the duration of one orbit which was, on the average,
$3-3.5$~ks.  As a result of this filtering, the total usable
observational time for Cyg~X-2 was more than 760~ks.

We constructed PDSs in the range $0.03125-128$\ Hz to analyze the
low-frequency ($<100$ Hz) variability of Cyg~X-2 and in the band
$128-2048$\ Hz to search for kHz QPO peaks.

No corrections were made for the background radiation and dead
time. We added a constant to the general model [see details in
\cite{vikhl}] to take into account the PCA deadtime effect, which
causes the overall level to be shifted to the negative region .

Fitting PDSs by a sum of constant and broken power law function did
not give acceptable results (according to the $\chi^2$ test).  The
transition between two slopes in PDS is rather smooth which results in
high residuals in the vicinity of the break frequency and large
resulting uncertainty in its measurement. A phenomenological model
which we use to fit the PDS continuum was more suitable
\begin{equation}
P(\nu)=A\nu^{-\alpha}[1+(\nu/\nu_{b})^{\beta}]^{-1}.
\label{pds_fit}
\end{equation}

For spectral analysis we accumulate the energy spectra from Standard2
PCA data mode using the FTOOLS software package \citep{bla95} and PCA
Energy spectra were also deadtime corrected according to "The RXTE
Cook Book''
\footnote{http://heasarc.gsfc.nasa.gov/docs/xte/recipes/cook\_book.html}.

We model the energy spectra using XSPEC astrophysical fitting package
in the energy range of 3-30 keV with an added systematic error of 1\%.
The modeling results produced an average reduced $\chi^{2}$ ($\chi^2$
divided by the degrees of freedom) $ \chi^{2}_{red} \sim 1$ across all
OBSIDs.  The spectra were modeled in XSPEC with a two-component
additive model consisting of a Comptonization (COMPTT model in XSPEC,
see Titarchuk 1994b; Hua \& Titarchuk 1995) and an Iron K$_\alpha$
line (Gaussian) model (Figure
\ref{PDSvsSpectra}, right panels). A value for the photon index
$\Gamma$ ($\Gamma=\alpha +1$) was inferred from the modeling results,
where $\alpha$ and $\gamma$, for spherical geometry of the Compton
cloud, are determined by
\begin{equation}
\alpha = -3/2 + \sqrt{9/4 + \gamma},
\label{alpha}
\end{equation}
\begin{equation}
\gamma = \pi^{2}m_ec^{2}/[3 (\tau + 2/3)^{2} kT ]
\label{Gamma}
\end{equation}
using the diffusion solution by \citet{ST80}.  The values for optical
depth $\tau$ and electron temperature $kT$ are determined from the
COMPTT model parameters and $m_ec^{2} = 511$ keV.  Error bars of
$\Gamma-$values were determined using the errors of $\tau$ and
$kT$. We present the results of our spectral analysis along with QPO
frequency for three representative observations in Table 1.

\section{Power spectrum and QPO identification}
\label{PDS}
\begin{figure}[tb]
\includegraphics[width=3.5in,height=3.5in,angle=0]{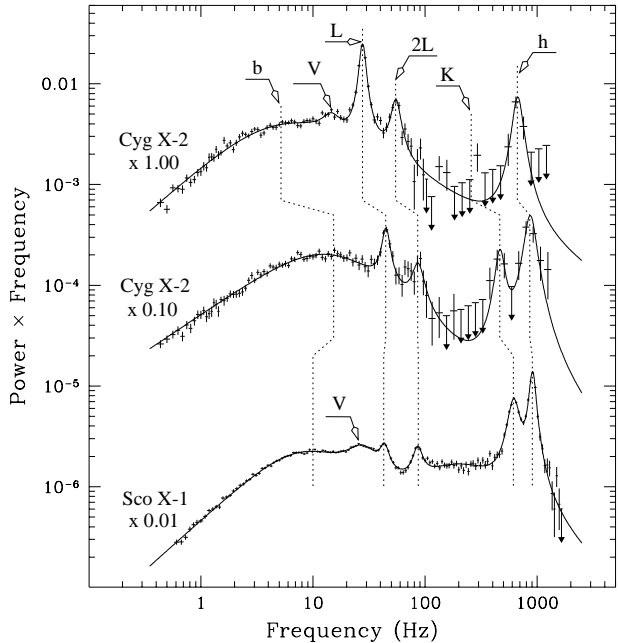}
\caption{Power Density Spectra of Z-sources Cyg~X-2 (as
a result of the presented study) and Sco~X-1 [adopted from \cite{tok}]
in units of $\nu\times P(\nu)$, (rms/mean)$^2$.  Upper limits
correspond to 1$\sigma$ confidence level.  Break frequency ($b$),
frequency of magnetoacoustic oscillations ($\nu_{V}$, HBO frequency
harmonics ($L$, $2L$), Keplerian ($K$) and hybrid ($h$) frequencies
are marked.  Similar features in all power density spectra are
connected by dotted lines.  Predicted value of Keplerian frequency in
the top spectrum is shown.  Lower peak in Sco~X-1 spectrum is shown as
a frequency of viscous oscillations $\nu_{V}$ (see text for details).
}
\label{CygX-2_ScoX-1}
\end{figure}

Figure \ref{PDSvsSpectra} illustrates the evolution of PDSs, observed
with RXTE/PCA on March 27, 1996 (filled circles), July 15, 1998
(starred data points), September 30, 2002 (open circles) and the
photon spectra.  The dash-dotted lines in Figure \ref{PDSvsSpectra}
depict a continuum fit to the HBO power spectra.  In Figure
\ref{CygX-2_ScoX-1} we show that the power spectra of Sco~X-1 and Cyg
X-2 are similar.

It is worth noting that the PDS's continuum power decreases as the
spectral state softens.  Strong HBO features are clearly seen in each
$\nu\times$ PDS diagram. HBO peaks of Cyg~X-2 PDSs vary in a wide
range from from 14~Hz to 60~Hz.  Both lower and upper kHz QPO peaks
are detected simultaneously only in an observation carried out on July
2, 1997 [see details of the data analysis and QPO identification in
\citet{kuz01}].  We obtain the high frequency QPO values $\nu_{\rm
K}=464.9\pm13.4$ Hz , $\nu_h=830.7\pm12.7$ Hz, with significances $3.4
\sigma$ and $5.8 \sigma$ for the QPO low and the high frequencies
respectively and the value of HBO peak, $\nu_L=44.70\pm0.35$ Hz, with
a significance of $7.7\sigma$.

Upper peak was rarely detected in observations corresponding to the
source position on HBO of Z-track.

\section{X-ray Spectra}

The sum of Comptonization and iron line components well describes the
Cyg~X-2 X-ray energy spectrum. The $N_{H}$ column was frozen at
Galactic value $0.22\times10^{22}$ cm$^{-2}$ [see \cite{dl90}].

Iron K$\alpha$ line energy was fixed at 6.4 keV.  The width of the
line was about 1 keV. The Gaussian equivalent width (EW) varies from
150 eV to 450 eV, showing a tendency to increase towards the soft
state, similar to BTK07 findings for Sco~X-1. However RXTE energy
resolution is poor at low energies and therefore the results for Iron
line parameters should be interpreted carefully.

The EW of the iron line can be also sensitive to the continuum model.
We obtain this result using COMPTT alone whereas other results in the
literature estimated this EW using much more complicated continuum
models. For example \cite{disalvo} reported on the results of a broad
band (0.1- 200 keV) spectral study of Cyg~X-2 using two Beppo-SAX
observations taken in 1996 and 1997 respectively.  These Beppo-SAX
spectra are fitted to a model consisting of a disk blackbody, a
Comptonization component, and two Gaussian emission lines at 1 keV and
6.6 keV, respectively. The addition of a hard power-law tail with
photon index 2, contributing 1.5 \% of the source luminosity.  The
second emission line, presumably K$_{\alpha}$ line at an energy of
6.6-6.7 keV has equivalent widths from 24 to 55 eV.  This value is
lower than that we found using only COMPTT component for the RXTE
data.  However our results in agreement with claims by \cite{disalvo},
\cite{kuu97} and \cite{smale93} that the addition of the iron K$_{\alpha}$
is required to fit the spectrum of Cyg~X-2.

\begin{figure}[tb]
\includegraphics[width=3.23in,height=3.23in,angle=0]{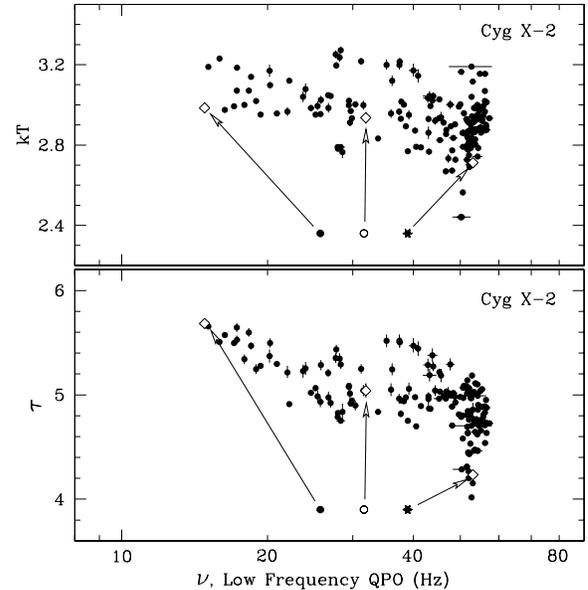}
\caption{
Best-fit Comptonization model parameters (xspec: COMPTT) versus
frequency of HBO peaks in Cyg~X-2. kT is the electron CC temperature
and $\tau$ is optical depth of CC cloud.  Open diamonds are the same
observational datasets as shown in Figure \ref{PDSvsSpectra}.  }
\label{kT_tau}
\end{figure}

For the COMPTT model we use spherical geometry.  We obtain the seed
photon temperature (T0) in the range 0.5-0.8 keV.  In Figure
\ref{kT_tau} we present the best-fit COMPTT parameters $\tau$ and $kT$
versus the observed HBO frequency.  We find that the Compton cloud
(CC) optical depth values varies in the range of $4-5.8$ as the CC
electron temperature varies in the range of $2.4-3.2$ keV.  It is
worth noting that these temperature values are very close to the color
temperature of blackbody emission of the NS compact photosphere (see
e.g. Titarchuk 1994a). It implies that the plasma is very close to
equilibrium with the photon field in the NS environment (CC).  On the
other hand high $\tau-$ values lead to the conclusion that any high
frequency oscillations ($\gtorder 400$ Hz) originating at the NS
surface must be suppressed by photon scattering in the Compton cloud
and are thus unlikely be detected [\citet {tit02}].  This statement is
probably true for all the Z-sources, for which high values of $\tau$
are usually found [see e.g. \citet{disalvo}, BTK07].

Although $\tau$ and $kT$ do not show clear correlations with the QPO
low (HBO) frequencies $\nu_{low}$ (see Fig. \ref{kT_tau}) there is a
well-established correlation between photon index $\Gamma$ and
$\nu_{low}$.
 
\section{Correlation of HBO  frequencies with photon index}

\begin{figure}[tb]
\includegraphics[width=3.22in,height=3.32in,angle=0]{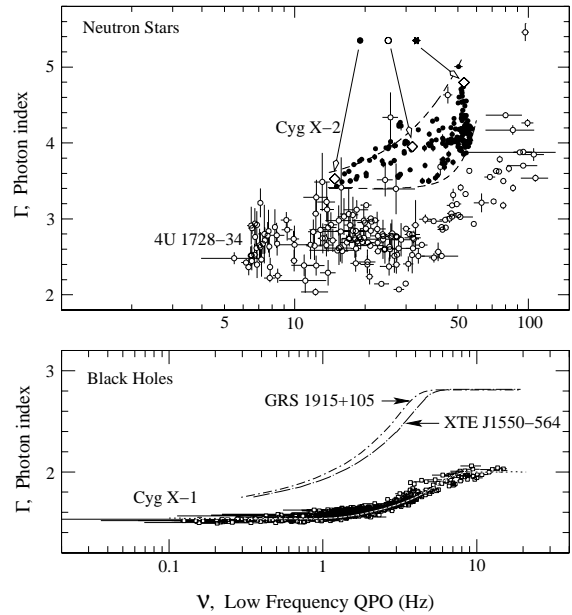}
\caption{Photon index $\Gamma$ 
as a basic characteristic of hardness of the energy spectrum vs low
frequency QPO of X-ray flux in the Neutron stars (top) and Black holes
(bottom) (see text).  {\it Top:} Z-source Cyg~X-2 (filled circles) and
Atoll-source 4U~1728-34 (open circles) observations demonstrate no
saturation effect at higher QPO frequencies. Dashed lines
schematically represent upper and lower ranges of $\Gamma$\ vs $\nu$\
variations in Cyg~X-2. Open diamonds are the same observational
datasets as shown in Figure~\ref{PDSvsSpectra}.  {\it Bottom:} Cyg~X-1
observational data points and best-fit model shapes to Cyg~X-1 (dotted
curve), GRS~1915+105 (short-dash-dotted curve) and XTE~J1550-564
(long-dash-dotted curve) data demonstrate correlation between photon
index $\Gamma$\ and Low-Frequency QPO [adopted from \cite{sha06}
]. Saturation effect is remarkable feature among black holes and is
observed when photon index reaches its maximum value at higher QPO
frequencies.  }
\label{index_QPO}
\end{figure}

Figure \ref{index_QPO} depicts the low-frequency (HBO) and $ \Gamma $
correlation revealed in Cyg~X-2 for which the Spearman Rank-Order
correlation coefficient $r_{sro}$ is 1 with the high precision
($r_{sro}=0.9998$).  Photon index $\Gamma$ was calculated using
$\tau$, $kT$ (presented in Fig. \ref{kT_tau}) and Eq.  (\ref{Gamma}).
A similar correlation of QPO vs $\Gamma$ in the high/soft state was
found in 4U~1728-34 by TS05.  The set of $\Gamma$, $\nu_{low}$ related
to the soft states of 4U~1728-34 and Cyg~X-2 are located in the right
upper corner of $\Gamma-\nu_{low}$ diagram.
   
In 4U~1728-34 the index-QPO correlation was found during evolution of
the source from low/hard state ($\Gamma\sim 2$) to soft state
($\Gamma\gax 4$). In contrast, Cyg~X-2 and Sco~X-1 (BTK07) always show
a soft spectrum and index-QPO correlation, as $\Gamma$ changes from
$3.3$ to $4$.  The Cyg~X-2 and Sco~X-1 photon indices $\Gamma=3.3-5$
are very close to that in the soft state of 4U~1728-34. However, in
both NS systems, inferred $\Gamma$ of the softest state ($4-5$) is
substantially higher than that of the soft state of black holes ($\sim
2.8$).
 
In the Z-source Cyg~X-2 (filled circles) and the Atoll-source
4U~1728-34 (open circles) observations demonstrate no saturation
effect at higher QPO frequencies. Dashed lines schematically represent
upper and lower ranges of $\Gamma$\ vs $\nu$\ variations in
Cyg~X-2. Open diamonds are the same observational datasets as shown in
Figure \ref{PDSvsSpectra}.  In Figure \ref{index_QPO} (bottom panel)
we present a correlation between photon index $\Gamma$\ and QPO low
frequency in Cyg~X-1, GRS~1915+105 (short-dash-dotted curve) and
XTE~J1550-564 (long-dash-dotted curve) [see details in
\cite{sha06}, hereafter ST06].

A saturation effect is a remarkable feature among black holes and it
is observed when photon index reaches its maximum value at higher QPO
frequencies.  The index saturation occurs when the total (disk plus
sub-Keplerian) mass accretion rate (in Eddington units, $\dot M_{\rm
Edd}=L_{\rm Edd}/c^2$) is more than 5 [see details in \cite{TZ98} and
\cite{tf04}, hereafter TF04].

This saturation effect in BHC sources, which is presumably due to
photon trapping in the converging flow, can be considered to be a BH
signature. We want to stress that this saturation is a
model-independent phenomenon found in the analysis of the RXTE
data. On the other hand, the saturation of the index with mass
accretion rate increase (which is strongly related to the QPO
frequency increase) has to apply to any BH because the photons are
unavoidably trapped in the accretion flow into BH.  In high/soft state
of BH the photon trapping effect in converging flow should be very
strong because the matter goes into a BH horizon and nothing comes
out. Photons are rather taken by the flow than they emerge.  It is a
necessary condition of BH presence in the accreting systems.

Thus, the index saturation with QPO frequency seen in the source
(rather than the presence of the extended hard tail in the soft state)
is a signature of the horizon. In fact, one can see high-energy tails
with indices at about the BH saturation value of 2.8 in high state
observations of weakly magnetized accreting NS binaries, for example,
GX~17+2 [\cite{disalvo00}, \cite{Fa05} and 4U~1728-34
(TS05)]. \cite{disalvo00} presented seven spectra of GX~17+2 observed
in 1999 by Beppo-SAX. They show this hard tail component in GX~17+2
gradually faded as the source moved toward the normal branch
(i.e. when the mass accretion rate in disk increases) where it was no
longer detectable.

TS05 analyzing RXTE archival data for 4U~1728-34 and we (in the
presented paper) analyzing RXTE data for Cyg~X-2 reveal the spectral
evolution of the Comptonized blackbody spectra and QPO frequencies.
Contrary to the BH sources, the indices of the 4U~1728-34 and Cyg~X-2
spectra do not saturate as the QPO frequency increases (see
Fig. \ref{index_QPO} for Cyg~X-2 case). They increase from 2 to 6 and
from 3.3 to 5 in the 4U~1728-34 and Cyg~X-2 correlations respectively
with no signature of saturation versus QPO frequency (or mass
accretion rate). The NS soft/state spectrum consists of blackbody
components that are only slightly Comptonized (the inferred photon
indices of the Comptonization Green's function are $\Gamma\gax
5$). Thus, one can claim (as expected from theory) that in NS sources
the thermal equilibrium is established for the high mass accretion
rate (soft) state. In BHs the equilibrium never established because of
the presence of the event horizon. The emergent BH spectrum, even in
the soft state, has a power-law component with an index that saturates
with mass accretion rate. It is worth noting that there is a
particular state in a BH source in which it can show signs of thermal
equilibrium: the emergent spectrum consists of one or two thermal
components. But no QPO is observed then. In this (very soft, thermal
dominated) state the source is presumably covered by a powerful wind
[see \cite{TSA07} hereafter TSA07] that thermalizes the radiation of
the central source and prevents us from seeing any QPO generated in
the source (see Fig. 6 in ST06).

We would like to emphasize once again that bulk inflow is present in a
BH when the mass accretion is high, but not in an NS, where the
presence of the firm surface leads to a high radiation pressure, which
eventually stops the accretion. The bulk inflow and all its spectral
features are absent in NS soft state; in particular, the saturation of
the index with respect to QPO frequency, which is directly related to
the optical depth and mass accretion rate, is observed in the soft
spectral state of only BHs and is therefore a particular signature of
a BH.

Thus the soft state index-QPO correlation can be a tool to determine
the nature of the compact object.  Due to the presence of a firm
surface, in NS sources photon field is approaching the equilibrium
with surrounding plasma as the source enters the high/soft state (when
$\Gamma$-value increases with QPO frequencies or with mass accretion
rate). In a BH source, however, solid surface is replaced by event
horizon, i.e. we have a drain instead of stopping boundary.

\section{Discussion}

In this paper we present the results of simultaneous timing and
spectral analysis of the NS Z-source Cyg~X-2.  The energy spectrum is
well described by the thermal Comptonization model (COMPTT). The only
additional component needed is the iron K$_{\alpha}$ line at $\sim6.4$
keV. This model fits all observations for all spectral states of Cyg
X-2.  The model parameters are the optical depth and temperature of
the Compton cloud which allows us to infer the photon index $\Gamma$
with rather small errors.

We found that our results for Cyg~X-2 are very similar to findings of
the BTK07 for Sco~X-1, where authors also successfully applied COMPTT
model to Sco~X-1 RXTE data [see also \cite{disalvo} for application of
COMPTT model to the broad energy band Beppo-SAX data from NSs].  This
strongly suggests that both Sco~X-1 and Cyg~X-2 form their spectra in
similar environments.  Additionally, the modeling results strongly
suggests that Sco~X-1 and Cyg~X-2 are always in the soft state, where
photon index $\Gamma$ is between 3.3 and 5.  Cooling of Compton cloud
is dictated by the strong soft emission of the disk and neutron star.

{\it High frequency pulsations originating at the NS surface cannot be
observed because of the CC optical depths in Sco~X-1 and Cyg~X-2, are
much greater than 1.} The NS high frequency pulsations are suppressed
because of photon scattering off cloud electrons.  This leads to a
significant conclusion that {\it the observed high (kHz) QPOs are
detected by the Earth observer because they originate at the outer
boundary of the CC}.  The NS emission, a $\sim 2.2$ keV blackbody, is
not seen at all in contrast to that of 4U~1728-34 (TS05).

One can ask a fair question why we do not see high frequency
pulsations in atoll-sources in the hard state.  In fact, the pulsated
NS emission is attenuated by scattering off electrons of the Compton
cloud and the detected signal consists of two components, the directed
and the scattered ones.  The directed component is exponentially
attenuated with attenuation factor $\exp{ (-T)}$ where $T$ is optical
path along the line of sight. Thus even the CC optical depth is
between 1 and 2 in the hard state of atoll-sources the optical path
along the line of sight can be a factor 2-3 more, i.e. $T=2-6$ and
consequently the directed component would be barely visible.  On the
other hand the pulsations of scattered component is completely is
washed if the CC optical depth more than two and the CC geometrical
thickness is order of few times of NS radius [see \cite{tit02}].

\cite{gag} recently tried to  revisit this Compton scattering scenario 
to explain lack of pulsations in the majority of LMXBs. They use
archival data of RXTE observations of three representative LMXBs
(GX~9+9, GX~9+1, and Sco~X-1) covering the full range of atoll/Z
phenomenology.  They argue that the optical depth of the Compton
region (corona) in these sources does not exceed $\tau\lax 0.2-0.5$
unless the electron temperature is very low, $kT_e <20$ keV.
Ultimately \cite{gag} conclude that lack of coherent pulsations cannot
be attributed to the electron scatterings because such small values of
the optical depth (that they inferred) are by far insufficient to
suppress the pulsations.

On the contrary if the electron temperature of the corona is low as
that is in our scenario (see Fig. 3, top ) then the Compton cloud is
optically thick [see Fig. 3, bottom, \cite{disalvo00}, \cite{gag}]. In
addition if the NS surface is surrounded by the quasi-spherical cloud
(which is presumably the case for the sources in the high/soft state)
then the NS pulsations are completely washed out.

The low frequency oscillations that are detected from Cyg~X-2 and Sco
X-1 can be also be considered as observational evidence of the
presence of the bounded Compton cloud in these sources (see details in
TSA07).

An other interesting example of the small optical depth interpretation
was presented by PSFK.  As we emphasize in Introduction section that
the real information in the 4U~0614+09 data is a detection of the
extended power laws which indices are about 2.4 but small optical
depth and high electron temperature are a matter of interpretation in
terms of the thermal Comptonization (TC) model.  In fact, BTK07 show
that the spectral index $\alpha=\Gamma-1$ is a reciprocal of the
Comptonization parameter $Y$ which is a product of average number of
scattering $N_{sc}$ and average fractional energy change per
scattering $\eta$. As for TC $\eta=4kT_e/m_ec^2$, i.e independent of
$\tau$ whereas for the converging (diverging) flow (CF) $\eta$ is
inverse proportional to $\tau$ (see e.g. Laurent \& Titarchuk 2007).
It implies that even when the CF $N_{sc}$ is large, the $Y$-parameter
(or $\Gamma$) can saturate to some low value because $N_{sc}$ is
proportional to $\tau$ in this case. In contrast, in the TC case one
needs to keep $\tau$ at small values in order to have values of $Y$
about $1/\alpha\approx1/1.4\approx0.7$ since the high energy cutoff
requires $kT_e\sim 200$ keV (see PSFK).  In fact, $N_{sc}\approx
Y/\eta$ and $\tau\approx Y/(2\eta)$ because $N_{sc}\approx2\tau$ in
the TC case.  Then one obtains a value of $\tau\approx
Y/(2\eta)\approx 0.7/3.2\approx0.2$ which is similar to that using the
TC model (see PSFK).
 
Titarchuk \& Farinelli 2007 (in preparation) made extensive
calculations of the hydrodynamics of the transition layer (Compton
cloud) presumably located between the neutron star surface and
geometrically thin Keplerian disk. They found that in the innermost
part of the transition layer (TL) of the size about NS radius the
matter is almost in free-fall regime.  Probably the hard spectra
detected from 4U~0614+09 are rather formed in this TL region than they
are a result of the thermal Comptonization in some optically thin
corona surrounding neutron star.  The real optical depth of the
Compton cloud (CC) can be large given that the 4U~0614+09 observations
show a relatively high temperature of the blackbody component of about
1.5 keV. Such a high temperature is definitely related to the NS
surface temperature and it can be only in the state when mass
accretion rate, and consequently CC optical depth, is high.  Thus, it
is not by chance, we do not see pulsations from 4U~0614+09 because
they are presumably blocked by optically thick Compton cloud.

\section{Conclusions}
We demonstrate a definite correlation of QPOs with photon index in Cyg
X-2 which has no sign of the index saturation at high values of QPO
frequency.  The similar correlations were previously reported in four
NSs: 4U~0614+09, 4U~1608-52, 4U~1728-34 and Sco~X-1.

The observed correlations of QPO frequencies with photon index in the
NS soft state lead us also to conclude that the Comptonization
efficiency is suppressed in the high/soft state of Cyg~X-2, namely the
disk and NS photons reach almost an equilibrium with surrounding
plasma of the Compton cloud.
 
In the BH high/soft state conversely, the index saturates with QPO
frequency (actually with mass accretion rate).  In terms of the
Physics of photon-plasma interaction this implies that photons are
trapped in converging flow i.e. they are driven by the flow across an
event horizon rather than emerging.  The bulk motion Comptonization
(upscattering) efficiency saturates with the mass accretion rate.

\section*{Acknowledgments}
We acknowledge the referee for his/her important questions and for the
constructive suggestions which improve of the paper quality. This
research has made use of data obtained through the High Energy
Astrophysics Science Archive Research Center Online Service, provided
by the NASA/Goddard Space Flight Center. SK was partially supported by
NASA Grant NAG5-12390.

\end{document}